\documentclass[a4paper]{article}

\RequirePackage[OT1]{fontenc}
\RequirePackage{amsthm}
\usepackage{graphicx}
\usepackage{natbib}
\usepackage{amssymb,amsthm}

\title{Self-consistent method for density estimation}

\author{Alberto Bernacchia\thanks{ Department of Neurobiology, Yale
    University, 333 Cedar Street, SHM-C400D, New Haven, Connecticut,
    alberto.bernacchia@yale.edu}$\ $ and Simone Pigolotti\thanks{
    The Niels Bohr International Academy, The Niels Bohr Institute,
    Blegdamsvej 17, DK-2100 Copenhagen, Denmark.}}

\begin{document}
\maketitle 
\begin{abstract}
  The estimation of a density profile from experimental data points is
  a challenging problem, usually tackled by plotting a
  histogram. Prior assumptions on the nature of the density, from its
  smoothness to the specification of its form, allow the design of
  more accurate estimation procedures, such as Maximum Likelihood.
  Our aim is to construct a procedure that makes no explicit
  assumptions, but still providing an accurate estimate of the
  density.  We introduce the self-consistent estimate: the power
  spectrum of a candidate density is given, and an estimation
  procedure is constructed on the assumption, to be released \emph{a
    posteriori}, that the candidate is correct. The self-consistent
  estimate is defined as a prior candidate density that precisely
  reproduces itself. Our main result is to derive the exact expression
  of the self-consistent estimate for any given dataset, and to study
  its properties. Applications of the method require neither priors on
  the form of the density nor the subjective choice of parameters. A
  cutoff frequency, akin to a bin size or a kernel bandwidth,
    emerges naturally from the derivation.  We apply the
  self-consistent estimate to artificial data generated from various
  distributions and show that it reaches the theoretical limit for the
  scaling of the square error with the dataset size.
\end{abstract}


\section{Introduction}

Every scientist has encountered the problem of estimating a continuous
density from a discrete set of data points. This may happen, for
example, when determining a probability distribution from a finite
Monte Carlo sample \citep{binder86}, rounding off the shape of a
galaxy from a collection of stars \citep{ripley90}, or assessing the
instantaneous firing rate of a neuron from a discrete set of action
potentials \citep{kass05}.  In all those cases, one can adopt two
different approaches: either assuming a given functional form for the
density \emph{a priori}, specified by a certain number of parameters,
or renouncing any prior knowledge (beyond that a density exists and,
in some cases, that is smooth).  These two approaches lead,
respectively, to parametric and non-parametric estimates.  We will
focus on the latter approach, although we will assume the knowledge of
the density as a reasoning tool, to be released \emph{a posteriori}.

The most popular non-parametric method is simply plotting a histogram,
but more sophisticated procedures have been developed. Kernel Density
Estimation (KDE) has been widely studied
\citep{silverman86,parzen61,wand95}: instead of counting the number of
points in separate bins, KDE constructs a smoothed picture of the data
as a superposition of kernel functions centered at the coordinates of
data points.  More formally, given a sample of $N$ data points (real
numbers), denoted by $\{X_j\}$ ($j=1,...,N$), the KDE estimate
$\hat{f}(x)$ is written as
\begin{equation}\label{eq:KDE}
\hat{f}_{KDE}(x)=\frac{1}{hN}\sum\limits_{j=1}^N K\left(\frac{x-X_j}{h}\right)  
\end{equation}
where $K(x)$ is the smoothing kernel and $h$ is the bandwidth.
Usually, the choice of $K(x)$ is not crucial \citep{silverman86}, while
$h$, which controls the degree of smoothing, has to be carefully
adjusted: the more concentrated data points are, the less smoothing is
necessary in order to obtain a good estimate of their density.  An
alternative non-parametric method is the Maximum Penalized
Likelihood (MPL, \citet{good71}), also known in the physics literature
as a regularization of Field Theory \citep{bialek96,holy97,schmidt00}:
it consists in performing a functional average of densities weighted
by their likelihood and by a measure of their smoothness.

In general, each non-parametric method depends on the arbitrary choice
of an adjustable parameter, such as the bin size in histograms, the
bandwidth in KDE, or the cutoff frequency in MPL and Field Theory.
Each of them regularizes the estimate and avoids overfitting of the
data points.  In most cases this corresponds to low-pass filtering,
i.e. cutting the high frequencies inherent to the discrete dataset,
and preventing the estimate from merely reproducing a narrow peak at
each data point. However, it would be desirable to devise methods
involving the least possible number of parameters, since their
determination usually involves some specific assumptions on the
distribution to be estimated (e.g. varying the cutoff parameter in MPL
and Field Theory precisely corresponds to different choices of the
Bayesian prior \citep{bialek96}).  Cross-validation techniques have
been previously applied for this purpose \citep{bowman84}, but they are
computationally expensive and have been seldom applied in the
literature.

In this study we show that a self-consistent approach leads to the
emergence of a natural cutoff frequency, and an estimate of the
density whose performance approaches the theoretical limit for the
scaling of the square error with the dataset size.  We start from the
observation, made in \citep{watson63}, that a unique "optimal"
convolution kernel can be derived as a function of the power spectrum
of the (unknown) density to be estimated. This result alone is of
little use, since the power spectrum of the true density is not known
\emph{a priori}.  However, in Section 2 we exploit the result by
defining the "self-consistent" estimate as the one whose associated
optimal kernel, applied to the sample dataset, returns the estimate
itself. Our main result is to derive the exact expression of the
self-consistent estimate for any given dataset, and to study its
properties.  In Section 3 we test the method on three different
problems: the estimates of Gaussian, Cauchy and Comb distribution.  In
all cases we show that the self-consistent estimate outperforms
existing methods and the mean integrated square error reaches the
optimal theoretical scaling $\sim N^{-1}$. Technical material is
  presented in the appendices: in Appendix 1 we replicate and extend
  the results on the existence of the optimal kernel. In Appendix 2 we
  provide details of the derivation of the self-consistent estimate.
  In Appendix 3 we prove that the self-consistent estimate converge
  almost surely to the true distribution for large $N$.

\section{The self-consistent estimate}

In this section, we define the self-consistent estimate, we derive its
exact expression, and we study its properties. We start from a
result derived by \citet{watson63}, that we replicate and extend in
Appendix 1.  The basic result is that a unique, optimal convolution
kernel can be derived as a function of the power spectrum of the
density to be estimated, where "optimal" is intended as minimizing the
mean integrated square difference between the true density and its
estimate.

Given a sample of $N$ data points (real numbers),
denoted by $\{X_j\}$ ($j=1\ldots N$), each independently drawn from a
probability density distribution $f(x)$, we write the
estimate as
\begin{equation}
\label{le}
\hat{f}(x)=\frac{1}{N}\sum_{j=1}^NK(x-X_j)
\end{equation}
where we assume $f, \hat{f}\in L^2$.  Note that (\ref{le}) does not
depend on any bandwidth $h$, contrary to the KDE estimate
(\ref{eq:KDE}).  Instead of choosing an arbitrary shape for the kernel
$K$, and looking for an optimal bandwidth (see
e.g. \citet{silverman86}), we rather look for an optimal shape of the
kernel.  It turns out that the Fourier transform $\kappa_{opt}(t)$ of
the optimal kernel $K_{opt}(x)$ is equal to (see Appendix 1 and
\citet{watson63})

\begin{equation}
  \label{ok}
  \kappa_{opt}(t)=\frac{N}{N-1+|\phi(t)|^{-2}}
\end{equation}
where $\phi(t)$ is the Fourier transform of the true density $f(x)$
(characteristic function).  The optimal kernel $K_{opt}(x)$ is
symmetric with respect to $x=0$, where it takes its maximum value.
Note that $|\phi(t)|$ in Eq.(\ref{ok}) requires the knowledge of the
true density, which is not available, hence Eq.(\ref{ok}) cannot be
used to compute the estimate in Eq.(\ref{le}) from the sample
observations $\{X_j\}$ alone.  We show in the following how to
circumvent this problem by a self-consistent approach.  Eq.(\ref{ok})
has been previously derived by \citet{watson63}, and has been used
for assessing the performance of specific kernels \citep{davis77}, as
well as for constructing blockwise estimators \citep{efromovich08}.

Although Eq.(\ref{ok}) cannot be used to compute the density estimate,
we make a step further and we write the Fourier transform
$\hat{\phi}(t)$ of the density estimate $\hat{f}(x)$ in Eq.(\ref{le}),
using the transformed kernel (\ref{ok}), as

\begin{equation}
\label{oe}
\hat{\phi}(t)=\Delta(t)\kappa_{opt}(t)=\frac{N\Delta(t)}{N-1+|\phi(t)|^{-2}}.
\end{equation}
where $\Delta(t)$ is the empirical characteristic function, i.e.

\begin{equation}
\label{deltafun}
\Delta(t)=\frac{1}{N}\sum_{j=1}^N e^{itX_j}
\end{equation}
 
Our approach is to construct an iterative procedure based on
Eq.(\ref{oe}), and to determine its exact fixed point.
We replace the unknown term $\phi$ in Eq.(\ref{oe}) with an initial
guess $\hat{\phi}_0$, and we denote the resulting estimate as
$\hat{\phi}_1$. Then, we try to obtain an improved estimate
$\hat{\phi}_2$ by using a kernel which is optimal for $\hat{\phi}_1$.
By iterating this procedure, we construct the following sequence of
estimates

\begin{equation}
\label{map}
\hat{\phi}_{n+1}=\frac{N\Delta}{N-1+|\hat{\phi}_n|^{-2}}.
\end{equation}
We search for a fixed point of the iteration, namely an estimate
$\hat{\phi}_{sc}$ for which

\begin{equation}
\label{eqsc}
\hat{\phi}_{sc}=\frac{N\Delta}{N-1+|\hat{\phi}_{sc}|^{-2}}
\end{equation}
This coincides with the density whose corresponding optimal kernel
applied to the data sample gives back the density itself.  We call the
resulting estimate a "self-consistent estimate".
We derive in Appendix 2 the stable solution of 
Eq.(\ref{eqsc}), which is equal to

\begin{equation}
\label{eq:scfin}
\hat{\phi}_{sc}(t)=\frac{N\Delta(t)}{2(N-1)}\left[1+\sqrt{1-\frac{4(N-1)}
{N^2|\Delta(t)|^2}}\right]I_A(t)
\end{equation}
where $I$ is the indicator function ($I_A(t)=1$ if $t\in A$,
$I_A(t)=0$ if $t\notin A$) and $A$ is the set of "accepted"
frequencies, i.e. the frequencies giving a nonzero contribution to the
estimate.  In order for (\ref{eq:scfin}) to be a stable solution of
(\ref{eqsc}), the set $A$ must be contained in $B$ ($A\subseteq B$),
where $t\in B$ if and only if

\begin{equation}
\label{thcond}
|\Delta(t)|^2\geq\frac{4(N-1)}{N^2}
\end{equation}
This condition sets a threshold for the amplitudes of 
frequencies $t$ below which $\hat{\phi}_{sc}(t)=0$.  
Hence, the contribution of small amplitude waves
is neglected, and this automatically determines the range of
frequencies to be considered for the estimate. In most practical
situations, the filter will cut the high frequency bands, but the
filter is not constrained to be low-pass, and it can rather select
different frequency bands.

\begin{figure}
\centering
\makebox{\includegraphics[scale=0.55]{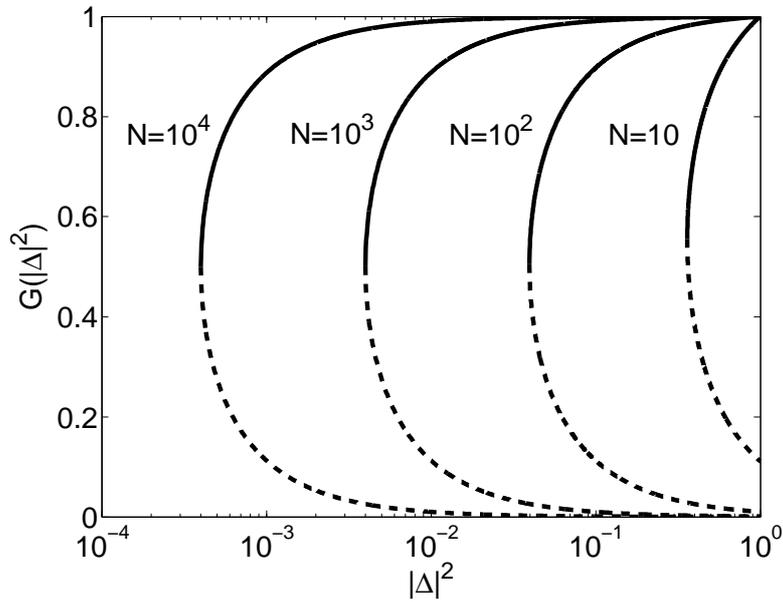}}
  \caption{Amplitude gain of the self-consistent estimate (full line),
    $G=|\hat{\phi}_{sc}|/|\Delta|$, as a function of the squared input
    amplitude $|\Delta|^2$ and for different values of the sample size
    $N$.  Dashed line shows the unstable solution (see Appendix 2).
    The amplitude gain exists only if the inequality (\ref{thcond}) is
    satisfied, and is always smaller than $1$, implying that the
    self-consistent estimate attenuates the amplitudes in the input.
    The exception is for $|\Delta|^2=1$, for which $G=1$,
    corresponding to the normalization condition.  $G$ tends to one
    for large values of $N$ (while the unstable solution vanishes),
    implying that the estimate tends to reproduce all frequencies of
    the input in that case.}
\label{gainex}
\end{figure}

The condition $A\subseteq B$ leaves the arbitrary choice of a subset
of frequencies among those above the threshold set by condition
(\ref{thcond}).  As demonstrated in Appendix 3, the self-consistent
estimate converges almost surely to the true density, provided that
$A$ is bounded, where the bound grows with $N$ (in addition, the
characteristic function is required to be integrable).  In practical
applications, a bounded interval must be necessarily implemented, and
we selected

\begin{equation}
\label{setA}
A=B\cap[-t^*,t^*]
\end{equation}
where, as defined above,
$B=\{t:|\Delta(t)|^2\geq\frac{4(N-1)}{N^2}\}$, and $t^*$ is set in
such a way that the inequality (\ref{thcond}) holds in one half of the
interval $[-t^*,t^*]$. The estimate is not sensitive 
to the choice of this interval (in applications, a $50\%$ change in the 
interval resulted, on average, in $1\%$ change in the estimate, see Section 3). 

Since $\hat{\phi}_{sc}$ and $A$ are bounded, the self-consistent estimate 
in Fourier space, Eq.(\ref{eq:scfin}), can be 
antitransformed back to the estimate in real space, i.e.

\begin{equation}
\label{eq:antisc}
\hat{f}_{sc}(x)=\frac{1}{2\pi}\int_{-\infty}^{+\infty} \, 
e^{-itx}\hat{\phi}_{sc}(t)dt
\end{equation}
Applications of the method are considered in the next section, 
here we describe its properties.
A graphical illustration of the filtering properties of the estimate
is given in Fig.\ref{gainex}, where the amplitude gain 
$G=|\hat{\phi}_{sc}|/|\Delta|$ is plotted.

The self-consistent estimate is normalized, i.e.
$\int_{-\infty}^{+\infty}\hat{f}_{sc}(x)dx=1$ or, equivalently,
$\hat{\phi}_{sc}(0)=1$, as a consequence of the normalization of the
empirical density ($\Delta(0)=1$).  Beside the zero frequency that is
kept intact (corresponding to the normalization condition), the
self-consistent estimate attenuates all other frequencies
($|\Delta|\leq1$ implies $|\hat{\phi}_{sc}|\leq|\Delta|$, see
Fig.\ref{gainex}).  Because $\hat{\phi}_{sc}(t)$ is continuous and
infinitely differentiable at $t=0$, all the moments of the
self-consistent estimate $\hat{f}_{sc}(x)$ exist.  The mean and
variance of $\hat{f}_{sc}(x)$ are equal to (see Appendix 2)

\begin{equation}
\label{ex}
E(x)=\frac{1}{N}\sum_{j=1}^NX_j
\end{equation}

\begin{equation}
\label{varx}
\mbox{Var}(x)=\frac{1}{N-2}\sum_{j=1}^N(X_j-E(x))^2
\end{equation}
While the mean is equal to the sample mean, the variance 
is larger than the sample variance as well as than its unbiased 
estimator, which is normalized by
$\frac{1}{N-1}$ instead of $\frac{1}{N-2}$.

\begin{figure}
\centering
\makebox{\includegraphics[scale=0.3]{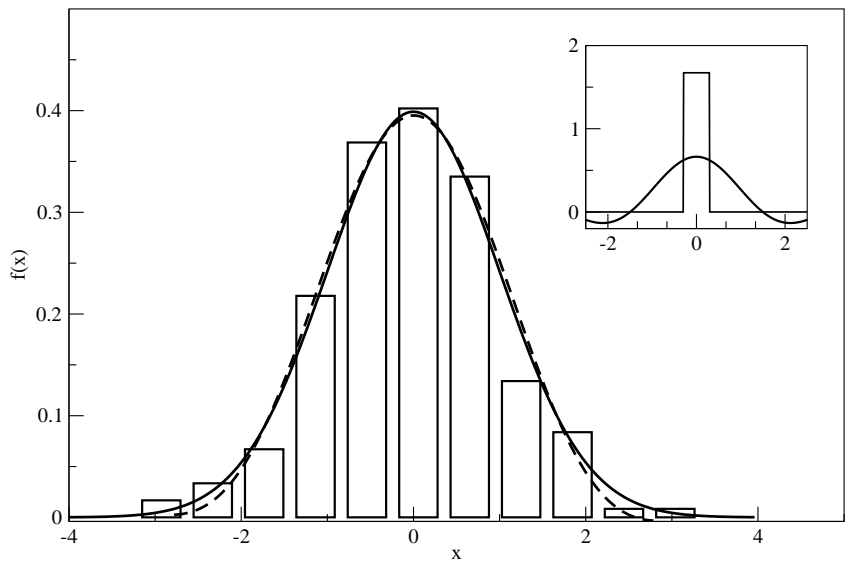}}
  \caption{Illustrative example of an estimate of a Gaussian density
    from $N=200$ sample points.  The true density is given by the full
    line. The dashed line is the self-consistent estimate, compared
    with a histogram having optimal binwidth. The inset shows the
    self-consistent kernel and the bin width of the histogram.
    \citep{scott1979}.}\label{figexample}
\end{figure}

A drawback of the self-consistent estimate is that it is not
guaranteed to be non-negative, while the true density is non-negative
(note that $|\hat{\phi}_{sc}(t)|^2\leq1$ holds from
Eq.(\ref{eq:scfin}), because $|\Delta|^2\leq1$, but that is a
necessary and not sufficient condition for $\hat{f}_{sc}(x)$ to be
non-negative).  
However, we seldom observed negative values in simulations. 
Those can be corrected without any error cost 
by translating the estimate
downward until the positive part is normalized to one, and setting to
zero the negative part \citep{glad03,ushakov99}.
In general, the restriction to a strictly non-negative
estimate has a cost, in terms of the mean integrated square error
$E(I)$, quantified by the decay exponent $\alpha$ of the error as a
function of the sample size, $E(I)\propto N^{-\alpha}$.  Among the
estimation procedures that properly give non-negative results,
histograms, MPL and Field Theories have $\alpha=2/3$
\citep{bialek96,holy97}, which is also the limit of KDE when the
density is discontinuous \citep{ushakov99}.  For a continuous density,
KDE improves to $\alpha=4/5$ \citep{silverman86}.  Applications of
estimates that are allowed to be negative reach better performance,
like "$m$-th order" kernels \citep{wand95, hall87, berlinet93}, that
have $\alpha=\frac{2m}{2m+1}$ (provided that the density is
$m$$-$$1$-th differentiable), while infinite order kernels
\citep{devroye92}, and the Sinc kernel \citep{davis77,schmidt00,glad07},
yield $\alpha=1$ (for infinitely differentiable densities, besides
logarithmic terms).  Hence, releasing the requirement of a
non-negative density estimate allows to improve the performance.
The optimal scaling $\alpha=1$ is also reached by parametric
estimators, such as Maximum Likelihood which are, however, strictly
non-negative.  We present below numerical results suggesting that the
self-consistent estimate (\ref{eq:scfin}) also reaches $\alpha=1$
for infinitely differentiable densities.

\section{Applications to artificial data}

\begin{figure}
\centering
\makebox{\includegraphics[scale=0.4]{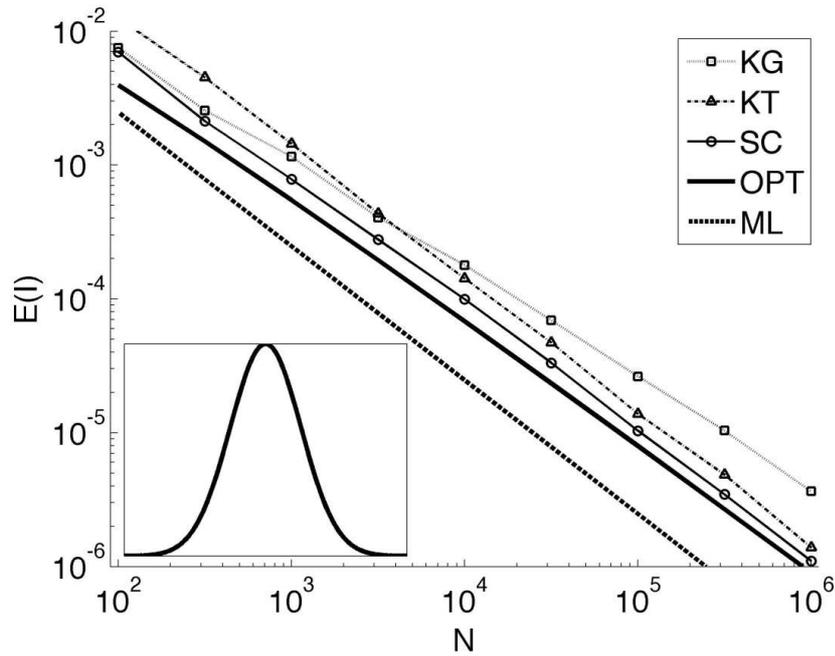}}
  \caption{Mean integrated square error $E(I)$ for the estimate of a
    Gaussian distribution (inset), as a function of the sample size
    $N$. Standard errors are about $5\%$ and smaller than
    the size of symbols. 
    Density estimates: Gaussian kernel (KG), Trapezoidal kernel (KT),
    self-consistent estimate (SC), each point is an average over 100
    realizations of the sample.  Theoretical bounds: optimal kernel
    (OPT), maximum likelihood (ML).  SC is applied without any prior
    knowledge, while ML assumes that the density is Gaussian and OPT
    requires its power spectrum in advance.}
    \label{figgaus}
\end{figure}

In this section we apply the self-consistent estimate to artificial
data.  The estimate $\hat{f}_{sc}(x)$ is constructed from a sample of
$N$ data points $\{X_j\}$ ($j=1\ldots N$), by using
Eqs.(\ref{deltafun}),(\ref{eq:scfin}-\ref{eq:antisc}).  As an
illustrative example, we present in Fig. \ref{figexample} the
application of the self-consistent estimate to a Gaussian sample of
$N=200$ points with respect to a histogram (where we chose the
optimal binwidth, see \citet{scott1979}).

We test more systematically the performance of the self-consistent
estimate on artificial datasets generated from three distributions:
Gaussian, Cauchy and Comb \citep{marron92}.  We compare the
performance of the self-consistent estimate (SC) with two kernel
estimators, the Gaussian kernel (KG, see \citet{silverman86}) and the
Trapezoidal kernel (KT, see \citet{politis03}), the latter being
representative of non strictly positive estimators.  For the Cauchy
distribution we also show the results for a kernel with a locally
adaptive bandwidth (APT, see \citet{silverman86, hossjer96}).  For
each density $f(x)$ and each estimator, we generate $100$ samples,
each composed of $N$ points randomly and independently drawn from
$f(x)$.  The performance is evaluated in terms of the mean integrated
square error,
Eq(\ref{Ifour}), where the mean is calculated over the $100$ sample
realizations.  We repeat the above procedure for different values of
$N$, ranging from $10^2$ to $10^6$.  In addition to the simulation
results, we also present the theoretical bound given by the optimal
kernel (\ref{ok}), i.e. the best an estimate of the type of
Eq.(\ref{le}) can achieve (OPT, see also Eq.(\ref{eq:minerr})).  In
the case of the Gaussian density we also show the theoretical bound
given by Maximum Likelihood (ML).  The bandwidths for the Gaussian and
Trapezoidal kernels are chosen from the sample data following
established empirical rules: $h=0.79*iq*N^{-1/5}$ for the Gaussian
kernel ($iq$ is the interquartile range, see Eq.(3.29) in
\citet{silverman86}), while for the Trapezoidal kernel
$h^{-1}=2\min\{m:|\Delta(m+s)|^2<c^2\log(N)/N,
\forall s\in(0,K_N)\}$, where we chose $K_N=\log(N)$ and we averaged
results over $2^{-2}\le c^2\le2^2$ (see \citet{politis03}).

First, we study the Gaussian distribution,
$f(x)=e^{-x^2/2}/\sqrt{2\pi}$.  Fig. \ref{figgaus} shows the mean
integrated square error $E(I)$ as a function of the sample size $N$,
for the self-consistent estimate (SC), the Gaussian and Trapezoidal kernels
(KG, KT), the optimal bound (OPT) and the Maximum Likelihood (ML).
For the Gaussian kernel with a fixed bandwidth $h$, the exact
expression for the error is

\begin{equation}
  E(I_{KG})=\frac{1}{2\sqrt{\pi}}\left(\frac{1}{Nh}-\frac{1}{N\sqrt{1+h^2}}
+1-\frac{2}{\sqrt{1+h^2/2}}+\frac{1}{\sqrt{1+h^2}}\right)
\end{equation}
The average value of the bandwidth is $h=1.06*N^{-1/5}$, which gives
an approximate value of the error $E(I_{KG})\simeq 0.33*N^{-4/5}$.
The error of the optimal estimate is given by Eq.(\ref{eq:minerr}),
and for a Gaussian density is equal to

\begin{equation}
E(I_{OPT})=\frac{\frac{N}{2\sqrt{\pi}(N-1)} 
\mbox{Li}_\frac{1}{2}(1-N)-1}{N-1}\simeq\frac{\sqrt{\log(N)}}{\pi N}
\end{equation}
where Li is the Polylogarithm function, defined by
Li$_s(z)=\sum_{k=1}^\infty \frac{z^k}{k^s}$.  The error for ML is
equal to

\begin{equation}
E(I_{ML})=\frac{7}{16\sqrt{\pi}}N^{-1}
\end{equation}

Fig.\ref{figgaus} shows that the error of the SC estimate is
  consistently smaller than both kernel estimates, KG and KT.  Both
SC and KT approach the theoretical OPT scaling $\sim
N^{-1}\sqrt{\log(N)}$ (see \citet{davis77}), while ML scales $\sim
N^{-1}$.  We stress that both ML and OPT require prior knowledge of
the density to be estimated. The former needs to know that the density
is Gaussian, the latter needs its spectrum in advance, while the
self-consistent method achieves the same scaling without any prior
assumption.

\begin{figure}[htb]
\centering
\makebox{\includegraphics[scale=0.15]{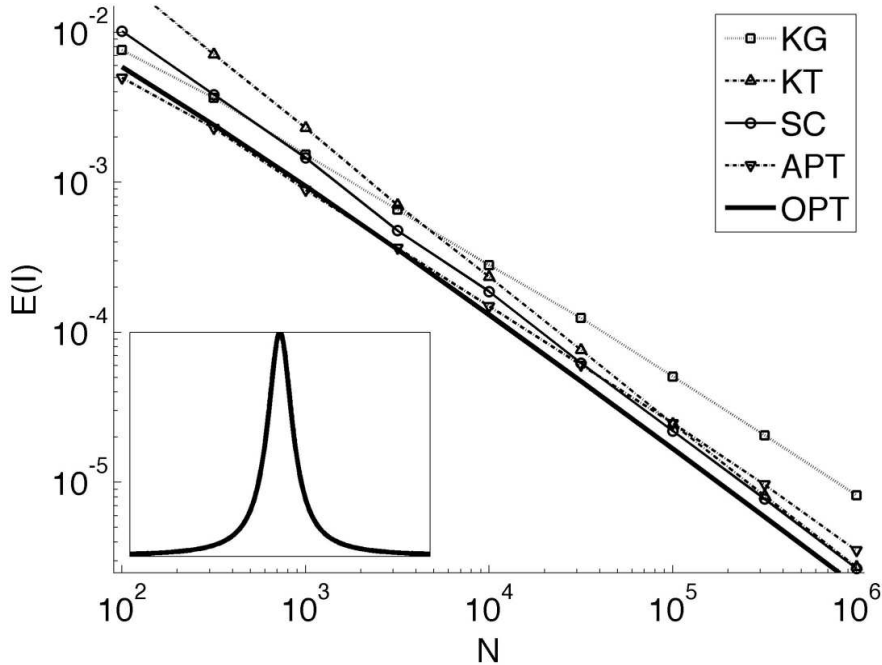}}
  \caption{Mean square error $E(I)$ for the estimate of a
    Cauchy distribution (inset) as a function of the sample size $N$.
    Standard errors are about $5\%$ and smaller than
    the size of symbols.
    Density estimates: Gaussian kernel (KG), Trapezoidal kernel (KT),
    adaptive kernel (APT), self-consistent estimate (SC), 
    each point is an average
    over 100 realizations of the sample.
    Theoretical bound: optimal kernel (OPT).
    SC is applied
    without any prior knowledge, APT applies to long-tailed distributions, OPT
    requires the power spectrum of the true density in advance.}
    \label{figcauchy}
\end{figure}

The second application is the estimate of a Cauchy distribution,
$f(x)=[\pi(1+x^2)]^{-1}$. The interest of this case comes from the
difficulties of binning long-tailed distributions, especially 
when the variance diverges. 
For the Gaussian kernel with a fixed bandwidth, the error is

\begin{equation}
E(I_{KG})=\frac{1}{2\sqrt{\pi}Nh}+\frac{1}{2\pi}+\frac{N-1}{2\sqrt{\pi}Nh}
e^{h^{-2}}\mbox{erfc}\left(h^{-1}\right)-\frac{\sqrt{2}}{\sqrt{\pi}h}e^{2h^{-2}}
\mbox{erfc}\left(\sqrt{2}h^{-1}\right)
\end{equation}

The average bandwidth is $h=1.58*N^{-1/5}$, for which $E(I_{KG})\simeq
0.55*N^{-4/5}$.  The error of the optimal estimate,
Eq.(\ref{eq:minerr}), in the case of the Cauchy distribution, yields

\begin{equation}
E(I_{OPT})=\frac{\frac{N}{2\pi (N-1)}\log(N)
-\frac{1}{2\pi}}{N-1}\simeq\frac{\log(N)}{2\pi N}
\end{equation}

In order to cope with long-tailed distributions, kernel methods have
been generalized to the "adaptive" kernel, which allows the bandwidth
to vary locally according to a first estimate of the density
\citep{silverman86, hossjer96}.  In Fig. \ref{figcauchy} we show the
mean square error as a function of $N$ for SC, KG, KT estimates, the
adaptive kernel estimate (APT), and the OPT bound.  For small
datasets, the adaptive kernel method performs best.  However, the
self-consistent method still shows better scaling with $N$, and its
error is lower for large sample sizes, with a crossover occurring for
$N$ between $10^4$ and $10^5$.  Again, the SC estimate performs better
than both kernel estimates KG and KT.  Both SC and KT estimates
approach the OPT scaling.
We stress that the adaptive method requires some prior knowledge: it
is used when one knows that the distribution is long-tailed and,
again, the optimal estimate requires prior knowledge of its power
spectrum.  Conversely, we applied the self-consistent method blindly,
in the same way as we did in the Gaussian case.

\begin{figure}[htb]
\centering
\makebox{\includegraphics[scale=0.15]{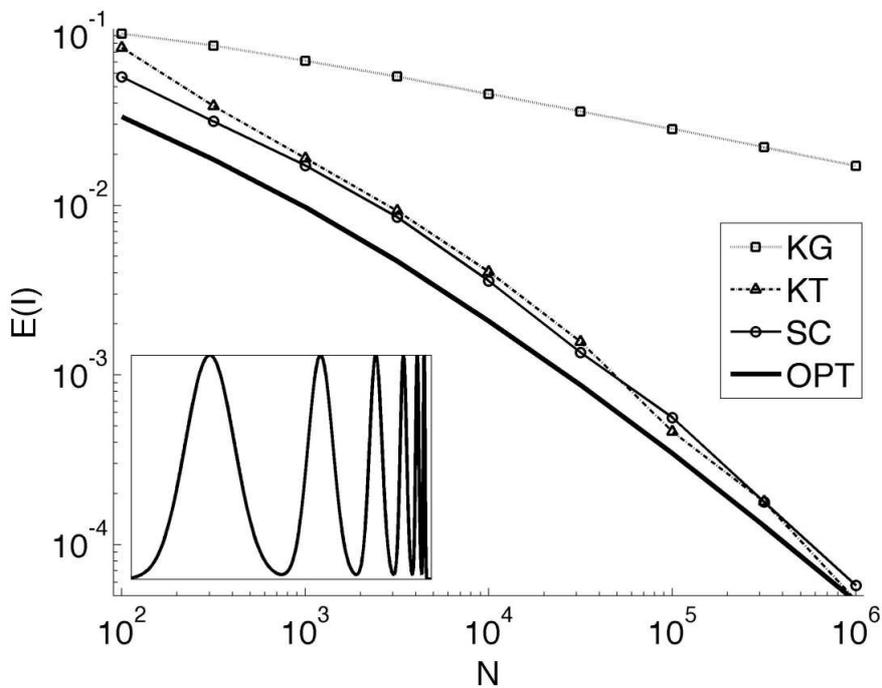}}
  \caption{Mean square error $E(I)$ for the estimate of a Comb
    distribution (inset) as a function of the sample size $N$.
    Standard errors are about $1\%$ and smaller than
    the size of symbols.   
    Density estimates: Gaussian kernel (KG), Trapezoidal kernel (KT),
    self-consistent estimate (SC), each point is an average over 100
    realizations of the sample.  Theoretical bound: optimal kernel
    (OPT).  SC is applied without any prior knowledge, OPT requires
    the power spectrum of the true density in advance.}
    \label{figcomb}
\end{figure}

The last application is the Comb distribution (\citet{marron92}),
which is multimodal, where different modes have different widths, and
its transform is affected by a large interval of frequencies.
Fig.\ref{figcomb} shows the results for SC, KG, KT estimates and the
OPT bound (which is computed numerically using Eq.(\ref{eq:minerr})).
Note that KG performs poorly in this case, because the empirical
bandwidth is unable to capture the large interval of frequencies of
the Comb distribution.  Instead, both SC and KT embrace an appropriate
interval of frequencies, estimated from the empirical characteristic
function, they perform much better than KG and approach the OPT
scaling.  

Finally, we remark that the self-consistent estimate may perform
poorly when applied to particular families of density functions.  As
explained in Appendix 3, if the density is not square integrable or
its characteristic function is not integrable, then the
self-consistent estimate is not guaranteed to converge to the true
density for large $N$.  Simulations (not shown) suggest that the
self-consistent estimate does not perform well when applied to the box
function (whose characteristic function is not integrable) and to the
$\chi^2$ distribution with one degree of freedom (which is not square
integrable).

\section{Discussion}

We presented a method that estimates a density in a self-consistent
way from a finite sample.  This approach produces a unique estimate
where no parameters have to be adjusted, and the only prior is the
belief in the self-consistent procedure.  On the other hand, the cases
in which there exists a widely accepted theoretical framework for the
system at hand would rather point to a Bayesian, parametric approach
in which a specific form for the density is postulated.

The self-consistent estimate converges to the true density for large
$N$ and it has a better performance and scaling than existing methods
for all the examples we studied. Together with the simplicity of
  its implementation, these features make it
  preferable for applications, especially when large datasets are available.
However, if the true density is not square integrable or its characteristic
function is not integrable, the self-consistent estimate is not guaranteed
to converge and it may perform poorly.

A frequency filter emerges naturally in the derivation
of the self-consistent estimate, as a function of the
data sample, and prevents overfitting of the data.  When long tails or
power-law behavior is suspected, one is tempted to use logarithmic
binning, but that has been shown to be highly inaccurate
\citep{clauset09,goldstein04}, and sometimes leading to the conclusion
that the density has power-law tails when it has not.  The
self-consistent method constitutes a good alternative when there are
no solid theoretical grounds to assume (or discard) a power law.

For small values of $N$, adaptive bandwidth kernels may perform better
than the self-consistent kernel (see Fig.\ref{figcauchy}).  Note that
if the bandwidth is allowed to vary locally, the performance of the
estimate is not bounded by the optimal kernel performance, since it
does not belong to estimates of the type of Eq.(\ref{le}), and can in
principle perform better.  Future studies will be devoted to extend
the self-consistent approach to estimates in which the kernel may vary
locally.  For example, a generic non-adaptive estimate can be
converted into an adaptive one by a transformation of $x$ leading to a
space with uniform measure \citep{ruppert94,periwal97}.  

The self-consistent method can be applied concretely to any problem in
which a density is sampled and there is no prior knowledge of its
functional shape.  For instance, the mode of the instantaneous firing
rate of a cortical neuron may indicate whether the choice of a
behaving monkey is triggered by a memorized object or a spatial cue
\citep{olson00}.  Another application could be the density estimate of
the spacings between zeros of the Riemann Zeta function, based on very
large numerical dataset, which could help investigate the related
mathematical conjectures \citep{odlyzko87}. Finally, the method can be
used to analyze samples obtained from Monte Carlo simulations
\citep{binder86}.

We remark that the method could not be applied to the case of 
integer numbers.
In general, when data points are uniformly spaced by multiples of any constant
length, the Fourier transform of their density is periodic. An amplitude 
threshold and a cutoff frequency would be meaningless in that case, 
and this would preclude any practical
application.  
However, when data points are integers there is usually
no need for filtering, and it is sufficient to use a histogram,
counting the occurrences of each number. 

Finally, a straightforward generalization is to consider a finite
interval $[a,b]$ instead of the entire line $(-\infty,+\infty)$, by
using Fourier series instead of Fourier transforms.  Another
possibility is to apply the method to high-dimensional distributions.
In our derivation, the relevant variables are scalar quantities, and
the $d-$dimensional analogy is obtained by using the $d-$dimensional
Fourier transform, i.e. by performing the integral
$\frac{1}{(2\pi)^d}\int_{\Re^d}dt^d$ instead of
$\frac{1}{2\pi}\int_{\Re}dt$.  Among many possible implementations,
the method could be then applied to the analysis of multielectrode
neuronal recordings \citep{brown04}, multivariate financial data
\citep{breymann03}, and reconstruction of Ramachandran angle
distributions \citep{kleywegt96}.

\section*{Appendix 1: Derivation of the optimal kernel}

In this section, we show that a unique "optimal" convolution kernel
can be derived as a function of the power spectrum of the density to
be estimated. A similar result has been presented in \citet{watson63}.
Given a sample of $N$ data points (real numbers),
denoted by $\{X_j\}$ ($j=1\ldots N$), each independently drawn from a
probability density distribution $f(x)$, we write the
estimate as
\begin{equation}
\label{linest}
\hat{f}(x)=\frac{1}{N}\sum_{j=1}^NK(x-X_j)
\end{equation}
The true density $f$ is assumed to be normalized, i.e.
$\int_{-\infty}^{+\infty}f(x)dx=1$.

We look for a kernel $K(x)$ such that the estimate (\ref{linest})
minimizes the mean integrated square error

\begin{equation}
\label{mse}
E(I)=E\int_{-\infty}^{+\infty}
\left[\hat{f}(x)-f(x)\right]^2dx
\end{equation}
where $E(\cdot)$ denotes an average
over all the possible realizations of the data sample $\{X_j\}$.  In
order to minimize (\ref{mse}), we follow a procedure for signal
deconvolution (the Wiener filter \citep{wiener49}).
We introduce the Fourier
transform of the unknown density (characteristic function)

\begin{equation}
\phi(t)=\int_{-\infty}^{+\infty}e^{itx}f(x)dx
\end{equation}
The normalization condition now reads $\phi(0)=1$.  We also call
$\hat{\phi}(t)$ and $\kappa(t)$ the Fourier transforms of the estimate
$\hat{f}(x)$ and of the kernel $K(x)$ respectively.  The mean
integrated square error (\ref{mse}) corresponds to the mean square
distance between the true density $f$ and the estimate $\hat{f}$, in
terms of the Euclidean metric in the Hilbert space $L^2$ (we assume
$f, \hat{f}, \phi, \hat{\phi}\in L^2$).  By means of Parseval's
theorem, we rewrite (\ref{mse}) in Fourier space as

\begin{equation}
\label{Ifour}
E(I)=\frac{1}{2\pi}E\int_{-\infty}^{+\infty}
\left|\hat{\phi}(t)-\phi(t)\right|^2dt
\end{equation}
It is straightforward to perform the average in Fourier space.
Applying the convolution theorem to Eq.(\ref{linest}), the transformed
estimate is equal to $\hat{\phi}(t)=\kappa(t)\Delta(t)$, where

\begin{equation}
\label{eq:delta}
\Delta(t)=\frac{1}{N}\sum_{j=1}^N e^{itX_j}
\end{equation}
is the empirical characteristic function. Note that $\Delta\notin L^2$ 
for any finite value of $N$, but
$\hat{\phi}\in L^2$ by assumption.  Using $E(\Delta)=\phi$,
and $E(|\Delta|^2)=|\phi|^2+N^{-1}\left(1-|\phi|^2\right)$,
we can rewrite the error as

\begin{equation}\label{IFourier}
E(I)=\frac{1}{2\pi}\int_{-\infty}^{+\infty}
\Big\{N^{-1}|\kappa|^2 \left(1-|\phi|^2\right)+|\phi|^2|1-\kappa|^2\Big\}dt.
\end{equation}
Since $f(x)$ is a density, it is normalized and non-negative, which
implies $|\phi|^2\leq 1$.  Then, the first term in the integral,
proportional to $1/N$, is non-negative: it corresponds to the error
due to the finite size of the sample, while the second term does not
depend on $N$.  These two sources of error are known as error variance
and error bias respectively \citep{silverman86}.  Among the possible
choices of the kernel, we search for the one minimizing the mean
integrated square error. Since Eq.(\ref{IFourier}) is quadratic in
$\kappa$, it is straightforward to find its global minimum, by setting
to zero the functional derivative of $E(I)$ with respect to $\kappa$,
i.e.

\begin{equation}
2\pi\frac{\delta E(I)}{\delta\kappa^*}=
N^{-1}\kappa\left(1-|\phi|^2\right)-|\phi|^2(1-\kappa)=0
\end{equation}
where the asterisk denotes complex conjugate.
This yields a unique, optimal kernel, which in Fourier space reads

\begin{equation}
  \label{eq:optkern}
  \kappa_{opt}(t)=\frac{N}{N-1+|\phi(t)|^{-2}}
\end{equation}
The optimal kernel satisfies the normalization condition
$\kappa_{opt}(0)=1$, because $\phi(0)=1$, and is a real function.
Since the density $f$ is real, then $|\phi(t)|=|\phi(-t)|$, which
implies that $\kappa_{opt}(t)$ is an even function.  Then its
antitransform $K_{opt}(x)$, the optimal kernel in the real space, is
also real and even and, because expression (\ref{eq:optkern}) is
non-negative, $K_{opt}(x)$ takes the maximum value at $x=0$, i.e. at
the coordinate of each data point in Eq.(\ref{linest}).

Using the expression for the transformed optimal kernel,
Eq.(\ref{eq:optkern}), we rewrite the estimate, Eq.(\ref{linest}), in
Fourier space, as

\begin{equation}
\label{optest}
\hat{\phi}(t)=\Delta(t)\frac{N}{N-1+|\phi(t)|^{-2}}.
\end{equation}
which we call the "optimal estimate".  The optimal estimate satisfies
the normalization condition, $\hat{\phi}(0)=1$, because $\Delta(0)=1$
and $\phi(0)=1$.  For infinite sample size ($N\rightarrow\infty$),
the optimal estimate reduces to the true density with probability one,
i.e. $\hat{\phi}(t)\rightarrow\phi(t)$, because $\Delta(t)\rightarrow\phi(t)$ 
(see \cite{csorgo83}, \cite{ushakov99} for the detailed sufficient conditions),
while the fractional term, the kernel, tends to one.
This is because an infinite sample would reproduce the true
density itself, without the need of any transformation of the data.
For finite $N$, the optimal estimate cuts the frequencies that have
less power in the true density, and hence are more subject to noise,
i.e. frequencies $t$ whose power is of the order
$|\phi(t)|^2\simeq1/N$ or less.

The above procedure is analogous to the derivation of the Wiener
filter for signal deconvolution \citep{wiener49}: the prior knowledge
of the power spectrum of both the signal, the unknown density, and the
noise establishes a unique criterion for optimal signal to noise
separation (note that, once the signal spectrum is given,
$|E(\Delta)|^2=|\phi|^2$, the assumption of independence of
data points allows the noise spectrum to be written as a function of
the signal, i.e. $E(|\Delta|^2)-|E(\Delta)|^2=N^{-1}(1-|\phi|^2)$).

We conclude this section by deriving the minimum square error obtained
by the application of the optimal kernel, that will be useful for
assessing the performance of practical applications of the method.  By
substituting the expression (\ref{eq:optkern}) of the optimal kernel
in (\ref{IFourier}), the associated minimum square error can be
written, after some algebra, as
\begin{equation}
  \label{eq:minerr}
\mbox{min}_K\,E(I)= \frac{K^{(N)}_{opt}(0)-K^{(1)}_{opt}(0)}{N-1}
\end{equation}
where we made explicit the dependence of the optimal kernel on the
sample size $N$, by writing $K_{opt}=K^{(N)}_{opt}(x)$.

Finally, note that $|\phi(t)|^2\leq1$ and $|\Delta(t)|^2\leq1$, which
implies from Eq.(\ref{optest}) that also the optimal estimate
satisfies $|\hat{\phi}(t)|^2\leq1$.  While this is a necessary
condition for the antitransform $\hat{f}(x)$ to be non-negative, it is
not sufficient, and $\hat{f}(x)$ is not guaranteed to be a
non-negative density.

\section*{Appendix 2: Derivation of the self-consistent estimate}

In this section we derive the expression for the self-consistent estimate 
$\hat{\phi}_{sc}$, Eq.(\ref{eq:scfin}), and we study its stability.
We start from Eq.(\ref{map}), the iterative map that we rewrite here

\begin{equation}
\label{itemap}
\hat{\phi}_{n+1}=\frac{N\Delta}{N-1+|\hat{\phi}_n|^{-2}}.
\end{equation}
We search for a fixed point of the iteration, namely $\hat{\phi}_{sc}$ such that

\begin{equation}
\label{sceq}
\hat{\phi}_{sc}=\frac{N\Delta}{N-1+|\hat{\phi}_{sc}|^{-2}}
\end{equation}
We derive in the following the two solutions (beyond the null solution) of
Eq.(\ref{sceq}) and we show that only one solution is stable with respect to the
iteration, Eq.(\ref{itemap}).
We start by taking the absolute value of Eq.(\ref{sceq}), in order
to obtain an equation for the single unknown variable $|\hat{\phi}_{sc}|$
(note that $\hat{\phi}_{sc}$ is complex valued).
Then, we multiply the expression by the denominator and by $|\hat{\phi}_{sc}|$
(leaving the null solution $\hat{\phi}_{sc}=0$), obtaining
a simple quadratic equation

\begin{equation}
\label{totti}
(N-1)\left|\hat{\phi}_{sc}\right|^2+1=N|\Delta|\left|\hat{\phi}_{sc}\right|
\end{equation}
Provided that $|\Delta|^2\geq\frac{4(N-1)}{N^2}$, this equation has
the following two solutions, denoted by the superscript $\pm$

\begin{equation}
\left|\hat{\phi}^{\pm}\right|=\frac{N|\Delta|}{2(N-1)}
\left[1\pm\sqrt{1-\frac{4(N-1)}{N^2|\Delta|^2}}\right]
\end{equation}
This solution gives the absolute value of $\hat{\phi}^\pm$.
By replacing this expression back into the right hand side of Eq.(\ref{sceq}),
we obtain the solution for $\hat{\phi}^\pm$, given by

\begin{equation}
\label{phipm}
  \hat{\phi}^{\pm}=\frac{N\Delta}{2(N-1)}
\left[1\pm\sqrt{1-\frac{4(N-1)}{N^2|\Delta|^2}}\right]
\end{equation}
While $\hat{\phi}^+$ is normalized, $\hat{\phi}^-$ is not, i.e.  when
$t=0$, $\Delta(0)=1$ implies $\hat{\phi}^+(0)=1$, while
$\hat{\phi}^-(0)=\frac{1}{N-1}$.  The two solutions are of very
different magnitudes: for large $N$ the solution $\hat{\phi}^-$
vanishes ($|\hat{\phi}^-|\simeq\frac{1}{N|\Delta|}$), while
$\hat{\phi}^+$ stays finite ($|\hat{\phi}^+|\simeq|\Delta|$).  For all
values of $|\Delta|$ for which the solutions $\hat{\phi}^{\pm}$ exist,
the following relation holds:
$|\hat{\phi}^+||\hat{\phi}^-|=\frac{1}{N-1}$.

We show that $\hat{\phi}^+$ is a stable solution of the iteration,
while $\hat{\phi}^-$ is unstable.  This can be seen by taking the
absolute value of Eq.(\ref{itemap}) and computing the derivative

\begin{equation}
\frac{d|\hat{\phi}_{n+1}|}{d|\hat{\phi}_n|}
\Bigg|_{|\hat{\phi}_n|=|\hat{\phi}^{\pm}|}=1\mp\sqrt{1-\frac{4(N-1)}{N^2|\Delta|^2}}
\end{equation}
This implies that, provided that the two solutions exist,
i.e. provided that $|\Delta|^2>\frac{4(N-1)}{N^2}$, then
$\hat{\phi}^+$ is stable (derivative smaller than one) and
$\hat{\phi}^-$ is unstable (derivative larger than one).  When
$|\Delta|^2=\frac{4(N-1)}{N^2}$, the two solutions annihilates in a
saddle node bifurcation.  For $|\Delta|^2<\frac{4(N-1)}{N^2}$ only the
null solution, $\hat{\phi}_{sc}$=0, is available.  That is always
stable, as can be checked by computing

\begin{equation}
\lim_{|\hat{\phi}_n|\rightarrow0}\frac{d|\hat{\phi}_{n+1}|}{d|\hat{\phi}_n|}=0
\end{equation}

In summary, when $|\Delta|^2\geq\frac{4(N-1)}{N^2}$, the iteration
(\ref{itemap}) reaches $\hat{\phi}_{sc}=0$ for
$|\hat{\phi}_0|<|\hat{\phi}^-|$ and $\hat{\phi}_{sc}=\hat{\phi}^+$ for
$|\hat{\phi}_0|\geq|\hat{\phi}^-|$, while for
$|\Delta|^2<\frac{4(N-1)}{N^2}$, the unique, globally stable solution
is $\hat{\phi}_{sc}=0$.  As described in the main text, we define the
set $B$ of $t$ values as

\begin{equation}
B=\left\{t:|\Delta(t)|^2\geq\frac{4(N-1)}{N^2}\right\}
\end{equation}
Hence, $\hat{\phi}_{sc}(t)=0$ when $t\notin B$.  However, $t\in B$
does not guarantee that $\hat{\phi}_{sc}(t)=\hat{\phi}^+(t)$, since a
small initial guess $|\hat{\phi}_0(t)|<|\hat{\phi}^-(t)|$ would
determine $\hat{\phi}_{sc}(t)=0$.  Then, we define $A$ as the set of
$t$ values for which $|\hat{\phi}_0(t)|\geq|\hat{\phi}^-(t)|$, and
hence $\hat{\phi}_{sc}(t)=\hat{\phi}^+(t)$.  The arbitrary choice of
the initial guess translates into the arbitrary choice of the set $A$,
provided that $A\subseteq B$, as required by the existence of the
nonzero solutions.  This is summarized in Eq.(\ref{eq:scfin}), and
concludes the derivation.

We conclude this section by calculating the mean and variance of the
self-consistent estimate $\hat{f}_{sc}$, Eq.(\ref{eq:antisc}). 
In a finite neighborhood of $t=0$, $\hat{\phi}_{sc}(t)=\hat{\phi}^+(t)$,
which is continuous and infinitely differentiable at $t=0$.
Then, the mean and variance
can be computed by the derivatives of $\hat{\phi}^+$ and
$|\hat{\phi}^+|^2$ at $t=0$, namely

\begin{equation}
\label{scm}
E(x)=-i\frac{d\hat{\phi}^+}{dt}\Bigg|_{t=0}=\frac{d\hat{\phi}^+}{d\Delta}\Bigg|_{\Delta=1}\left(-i\frac{d\Delta}{dt}\Bigg|_{t=0}\right)
\end{equation}  

\begin{equation}
\label{scv}
\mbox{Var}(x)=-\frac{1}{2}\frac{d^2|\hat{\phi}^+|^2}{dt^2}\Bigg|_{t=0}=\frac{d|\hat{\phi}^+|^2}{d|\Delta|^2}\Bigg|_{|\Delta|^2=1}\left(-\frac{1}{2}\frac{d^2|\Delta|^2}{dt^2}\Bigg|_{t=0}\right)
\end{equation}
where we used the chain rule and the fact that $|\Delta|^2$ is an even
function, hence $\frac{d|\Delta|^2}{dt}\big|_{t=0}=0$.  It is
straightforward to show that the terms in round brackets in
Eqs.(\ref{scm},\ref{scv}) are, respectively, equal to the sample mean
and sample variance.  Because $|\Delta|$ is even, then
$\frac{d\hat{\phi}^+}{d\Delta}\big|_{\Delta=1}=1$ and, by
differentiating Eq.(\ref{totti}), we obtain
$\frac{d|\hat{\phi}^+|^2}{d|\Delta|^2}\big|_{|\Delta|^2=1}=N/(N-2)$.
Finally, Eqs.(\ref{scm},\ref{scv}) can be rewritten as

\begin{equation}
E(x)=\frac{1}{N}\sum_{j=1}^NX_j
\end{equation}

\begin{equation}
\mbox{Var}(x)=\frac{1}{N-2}\sum_{j=1}^N(X_j-E(x))^2
\end{equation}

\section*{Appendix 3: Asymptotic convergence of $\hat{f}_{sc}$}
In this section we investigate the asymptotic (large $N$) behavior of
the self-consistent estimate.  We study the sufficient conditions for
the estimate $\hat{f}_{sc}$ to converge to the true density $f$ for
$N\rightarrow\infty$.  In particular, we prove the following

\subsection*{{\emph  Theorem}}

If the true density $f(x)$ is square integrable and its transform
is integrable,
then the self-consistent density
estimate $\hat{f}_{sc}(x)$, defined by 
Eqs.(\ref{deltafun}),(\ref{eq:scfin}-\ref{eq:antisc}),
converges almost surely to the true density for large $N$,
under the additional assumptions

\begin{equation}
\label{tstar1}
\lim_{N\rightarrow\infty} t^*=\infty
\end{equation}

\begin{equation}
\label{tstar2}
\lim_{N\rightarrow\infty} \frac{t^*}{\sqrt{N}}=0
\end{equation}

\subsection*{{\emph Proof}}

Because the true density $f(x)$ and the self-consistent 
estimate $\hat{f}_{sc}(x)$ are both square integrable, 
we can express them as Fourier transform of, 
respectively, $\phi(t)$ and $\hat{\phi}_{sc}(t)$.
By assumption, the characteristic function is integrable, i.e.

\begin{equation}
\label{integrability}
\int|\phi(t)|dt<\infty
\end{equation}

In the following sequence of inequalities, we find an upper
bound for the difference between the true density and
its estimate, and we use the fact that $\hat{\phi}_{sc}(t)=0$ for $|t|>t^*$.
In order to prove the theorem, we show that the upper bound
tends to zero for large $N$.

\begin{eqnarray*}
\left|\hat{f}_{sc}(x)-f(x)\right|=\left|\frac{1}{2\pi}\int_{-\infty}^{+\infty}e^{-itx}\left[\hat{\phi}_{sc}(t)-\phi(t)\right]dt\right|\leq\\
\leq\frac{1}{2\pi}\int_{-\infty}^{+\infty}\left|e^{-itx}\right|\left|\hat{\phi}_{sc}(t)-\phi(t)\right|dt=
\frac{1}{2\pi}\int_{-\infty}^{+\infty}\left|\hat{\phi}_{sc}(t)-\phi(t)\right|dt=\\
=\frac{1}{2\pi}\int_{-t^*}^{t^*}\left|\hat{\phi}_{sc}(t)-\phi(t)\right|dt+\frac{1}{2\pi}\int_{|t|>t^*}\left|\phi(t)\right|dt=\\
=\frac{1}{2\pi}\int_{-t^*}^{t^*}\left|\hat{\phi}_{sc}(t)-\Delta(t)+\Delta(t)-\phi(t)\right|dt+
\frac{1}{2\pi}\int_{|t|>t^*}\left|\phi(t)\right|dt\leq\\
\leq \frac{1}{2\pi}\int_{-t^*}^{t^*}\left|\hat{\phi}_{sc}(t)-\Delta(t)\right|dt+
\frac{1}{2\pi}\int_{-t^*}^{t^*}
\left|\Delta(t)-\phi(t)\right|dt+\frac{1}{2\pi}\int_{|t|>t^*}\left|\phi(t)\right|dt
\end{eqnarray*}
where $\Delta(t)$ is the empirical characteristic function, see Eq.(\ref{deltafun}).
For increasing $N$, $t^*$ increases following the limiting bounds given
by Eqs.(\ref{tstar1},\ref{tstar2}).
Then, the second integral in the last expression tends to zero 
because of theorem 1 in ref. \citep{csorgo83}, 
while the third integral tends to zero because of the 
integrability of the characteristic function $\phi(t)$,
Eq.(\ref{integrability}).
In order to prove the theorem, we have to demonstrate that
also the first integral tends to zero for large $N$.
We use the expression of $\hat{\phi}_{sc}$, 
Eq.(\ref{eq:scfin}), and we denote by $\Delta^+$
and $\Delta^-$ the set of values of $t$ for which
$|\Delta(t)|^2$ is, respectively, above or below
the threshold set by Eq.(\ref{thcond}),
i.e. $|\Delta(t)|^2\geq 4(N-1)/N^2$ or
$|\Delta(t)|^2<4(N-1)/N^2$.
Then, the first integral is equal to

\begin{eqnarray*}
\frac{1}{2\pi}\int_{(-t^*,t^*)\cap\Delta^+}\left|\Delta(t)\right|\left\{1-
\frac{N}{2(N-1)}\left[1+\sqrt{1-\frac{4(N-1)}{N^2|\Delta(t)|^2}}\right]
\right\}dt+\\
+\frac{1}{2\pi}\int_{(-t^*,t^*)\cap\Delta^-}\left|\Delta(t)\right|dt
\end{eqnarray*}
where the term in curly brackets is non-negative.
The first term can be expanded by means of 
the inequality $\sqrt{1-x}\geq 1-\sqrt{x}$
for $x\in(0,1)$, while in the second term we
substitute the integrand with its maximum
value in the interval, and we use the fact that
the length of the interval is smaller than $2t^*$.
Then, the above expression is smaller than

\begin{eqnarray*}
\leq \frac{1}{2\pi}\int_{(-t^*,t^*)\cap\Delta^+}\left[
\frac{1}{\sqrt{N-1}}-\frac{\left|\Delta(t)\right|}
{N-1}\right]dt+
\frac{2t^*\sqrt{N-1}}{\pi N}\leq\\
\leq \frac{1}{2\pi}\int_{(-t^*,t^*)\cap\Delta^+}\left[
\frac{1}{\sqrt{N-1}}+\frac{\left|\Delta(t)\right|}
{N-1}\right]dt+
\frac{2t^*\sqrt{N-1}}{\pi N}\leq\\
\leq
\frac{t^*}{\pi}\left[\frac{1}{\sqrt{N-1}}+
\frac{1}{N-1}+\frac{2\sqrt{N-1}}{N}\right]
\end{eqnarray*}
where we used $|\Delta(t)|\leq 1$ in the last inequality.
Because of Eq.(\ref{tstar2}),
the last expression tends to zero for large
$N$, thus proving the theorem.

\subsection*{{\emph Comment}}
The above theorem assumes that the true density is square integrable
and its transform integrable.  We expect the self-consistent estimate
to perform poorly when applied to densities that do not meet those
criteria, and its asymptotic consistency is not guaranteed.  Examples
of such distributions include the box function and the $\chi^2$
distribution with one degree of freedom.

The condition (\ref{tstar2}) gives the limit on the frequency bound
$t^*$, setting the maximum increase of $t^*$ with $N$.  In
simulations, the magnitude of the bound $t^*$ depends on the threshold
of the empirical characteristic function, Eq.(\ref{thcond}), in such a
way that its amplitude is larger than the threshold in half of the
interval $(-t^*,t^*)$.  Now we argue that this choice satisfies the
assumptions of the theorem, without giving a formal derivation.

To simplify the argument, we assume that the distribution is
sufficiently smooth so that the following limit exists
\begin{equation}
\label{linking}
\lim_{|t|\rightarrow\infty}t|\phi(t)|=0,
\end{equation}
notice that, if the above limit exist, then it must be equal to zero as
a consequence of the integrability of the characteristic function,
Eq.(\ref{integrability}). We also approximate the empirical
characteristic function $\Delta(t)$ with the true characteristic
function $\phi(t)$, which is reasonable provided that $N$ is large and
$t$ does not increases exponentially with $N$ (see \citet{csorgo83}).
Then, the threshold condition (\ref{thcond}) becomes
$|\phi(t)|^2\geq4(N-1)/N^2$.  Hence, $t^*$ increases in such a way
that, for large $N$, the order of magnitude of the characteristic
function is $|\phi(t^*)|\sim1/\sqrt{N}$.  Substituting this expression
in Eq.(\ref{linking}) we get

\begin{equation}
\lim_{t^*\rightarrow\infty}t^*|\phi(t^*)|=
\lim_{N\rightarrow\infty}\frac{t^*}{\sqrt{N}}=0
\end{equation}
which correspond to condition (\ref{tstar2}).
On the other hand, since the true cumulative distribution
is continuous, then $|\phi(t)|>0$ for any finite value of $t$,
and this implies that $t^*$ tends to infinity for
large $N$, as required by the condition (\ref{tstar1}).

In conclusion, the above theorem gives the sufficient 
conditions for the asymptotic consistency of the estimate,
especially concerning the finite interval of frequencies
set by the bound $t^*$.
From the above arguments, we expect the recipe for 
$t^*$ used in simulations to guarantee the
asymptotic convergence of the estimate.

\section*{Acknowledgements}

AB would like to thank Alfonso Sutera for introducing him to the
issues of nonparametric density estimation. The authors further thank
Yali Amit, Massimo Cencini, Rishidev Chaudhuri, Andrew Jackson, John Murray,
Umberto Picchini and Angelo Vulpiani for their useful comments on a
preliminary version of the manuscript.

\end{document}